\documentclass{Interspeech}

\interspeechcameraready

\title{Meta-PerSER: Few-Shot Listener Personalized Speech Emotion Recognition via Meta-learning}

\usepackage[symbol*]{footmisc}
\setfnsymbol{wiley}
\author[affiliation={1}]{Liang-Yeh}{Shen$^*$}
\author[affiliation={1}]{Shi-Xin}{Fang$^*$}

\author[affiliation={1}]{Yi-Cheng}{Lin}
\author[affiliation={2}]{Huang-Cheng}{Chou}
\author[affiliation={1}]{Hung-yi}{Lee}

\affiliation{National Taiwan University}{Taiwan}

\affiliation{}{Independent Researcher}{Taiwan}
\email{\{b10901005, b10507008, f12942075, hungyilee\}@ntu.edu.tw, huangchengchou@gmail.com}

\usepackage{comment}
\usepackage{pgfplots}
\pgfplotsset{compat=1.18}  % Adjust compatibility if necessary
\usepackage{tikz}
\usepackage{booktabs}
\usepackage{multirow}
\usepackage{colortbl}

\begin{document}

\maketitle

\begingroup
  \renewcommand\thefootnote{*}
  \footnotetext{Equal contribution}
\endgroup


\begin{abstract}

This paper introduces \textbf{Meta-PerSER}, a novel meta-learning framework that personalizes Speech Emotion Recognition (SER) by adapting to each listener’s unique way of interpreting emotion. Conventional SER systems rely on aggregated annotations, which often overlook individual subtleties and lead to inconsistent predictions. In contrast, Meta-PerSER leverages a Model-Agnostic Meta-Learning (MAML) approach enhanced with Combined-Set Meta-Training, Derivative Annealing, and per-layer per-step learning rates, enabling rapid adaptation with only a few labeled examples. By integrating robust representations from pre-trained self-supervised models, our framework first captures general emotional cues and then fine-tunes itself to personal annotation styles. Experiments on the IEMOCAP corpus demonstrate that Meta-PerSER significantly outperforms baseline methods in both seen and unseen data scenarios, highlighting its promise for personalized emotion recognition.

\end{abstract}

\section{Introduction}

\emph{Speech Emotion Recognition} (SER) aims to automatically identify emotional states from vocal cues \cite{schuller2004speech, khalil2019speech, nagase2021speech}. Despite substantial progress with deep learning, SER remains challenging due to the complexity of emotional expressions and the subjectivity of emotion perception \cite{ren2024emo}. One fundamental challenge in SER is the variability across speakers and listeners \cite{Mower_2009,Sethu_2019}. 

Traditional SER models typically assume a one-size-fits-all approach, learning a global mapping from acoustic features to emotion labels. However, this approach neglects individual differences: speakers have unique vocal characteristics, and annotators (listeners) often interpret the same speech differently, leading to inconsistent labels and noisy emotion datasets. Although speaker normalization or adaptation techniques \cite{moine2021speaker, tran2023personalized} address speaker variability, they still strive for a universal model that might not account for each listener's perception of emotion. In practice, a model that could adapt to each individual's emotion perception—essentially a listener-personalized SER system—would be highly desirable, yet existing SER approaches lack this personalization capability.

A key challenge in listener-personalized SER is the high variability in emotion annotations across individuals. Aggregating annotations from multiple listeners into a single ground-truth label risks losing personal perception and introducing biases \cite{plank2022problem, aroyo2023dices, prabhakaran2021releasing, Wu_2024}. Conversely, training separate models for each listener requires substantial labeled data and computational resources, which is impractical for real-world deployment \cite{Chou_2019,davani2022dealing}. Additionally, real-world applications require quick adaptation to new, unseen listeners with minimal labeled data, a scenario where traditional SER approaches struggle \cite{hospedales2021meta}.

Meta-learning offers a promising avenue to tackle the aforementioned challenges by enabling models to learn from a distribution of tasks \cite{finn2017model,ravi2017optimization}. Such algorithms can rapidly adapt to new tasks with only a few training examples, making them well-suited for data-scarce scenarios. In SER, each task could correspond to a specific condition (e.g., a new speaker or a listener's annotation style). By training on many such tasks, the model learns how to quickly adapt its parameters to a new task. Recent studies have started exploring meta-learning for SER to improve cross-corpus performance or reduce annotation noise \cite{fujioka20_interspeech, metaser, ottoni2023deep, LIU2023110766, gandhi2023efficacy}, but to our knowledge, none have focused on personalizing the model to individual listeners' interpretations. While the study \cite{Chou_2020} builds personalized SER models, it focuses on dimensional SER (e.g., arousal and valence), whereas our work focuses on categorical SER (e.g., happy or angry).

In this paper, we propose \textit{Meta-PerSER}, a novel meta-learning framework for SER that incorporates the listener's perspective into the learning process to achieve listener-personalized adaptation. During meta-training, the framework simulates adapting to different listeners by treating each listener's labeled data as a separate task. This approach allows the model to learn a good initialization that can be efficiently fine-tuned for a new listener using only a small amount of that listener's data. As a result, our method can quickly calibrate to a new listener's subjective emotion perception, thus addressing label ambiguity by aligning predictions with each listener's expectations. Unlike traditional SER models that yield a fixed classifier, Meta-PerSER is inherently adaptable, bridging the gap between generic and personalized emotion recognition. We will release our code\footnote{https://github.com/Jeffabcd/Meta-PerSER}. 

\begin{figure*}[t]  
    \centering
    \includegraphics[width=\textwidth]{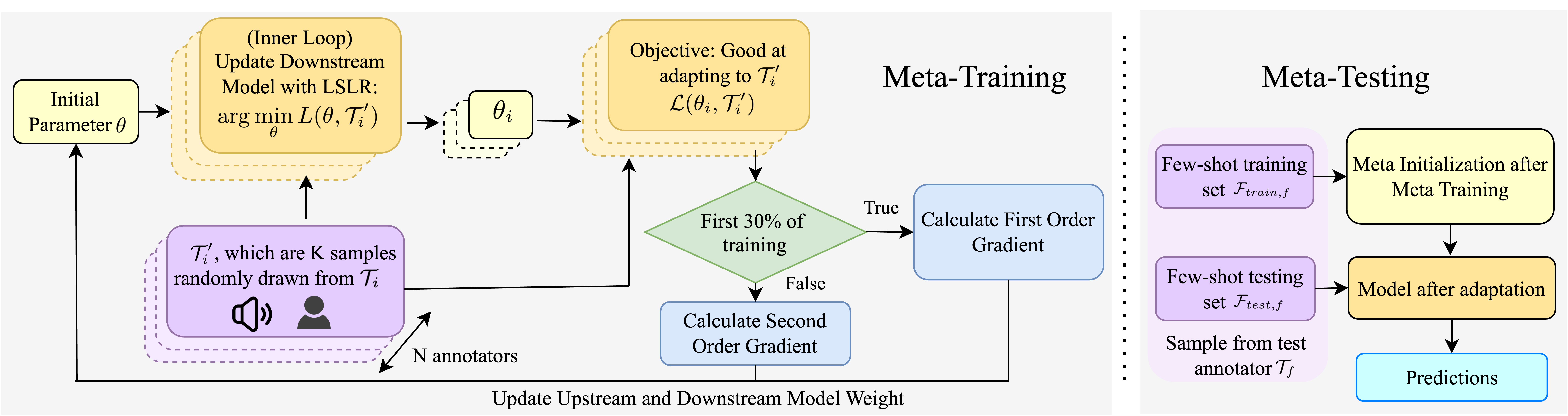} 
    \vspace{-6mm}
    \caption{\small Framework of proposed Meta-PerSER.}
    \vspace{-6mm}
    \label{fig:meta_perser} 
\end{figure*}

Experiments on a well-known emotion database, the IEMOCAP \cite{busso2008iemocap}, show that Meta-PerSER outperforms baselines, especially in few-shot settings with limited new-listener data. This underscores the benefit of personalization in SER. In summary, our main contributions are:
\begin{itemize} 
    \item \textbf{Listener-personalized SER via meta-learning:} We introduce the first SER framework that personalizes to individual listeners' categorical emotion interpretations. 
    \item \textbf{Efficient adaptation with limited data:} We develop a meta-training strategy that rapidly adapts to a new listener using only a small amount of data. 
    \item \textbf{Improved performance and robustness:} We empirically demonstrate that Meta-PerSER improves emotion recognition accuracy and robustness to annotation differences, outperforming state-of-the-art baselines.
\end{itemize}

\section{Methodology}
\label{s:method}

\subsection{Backbone SER Framework}
We employ a unified model architecture based on the s3prl toolkit \cite{Yang_2021}. Our SER framework comprises two principal components: an upstream module and a downstream module. The upstream module leverages pre-trained self-supervised learning (SSL) models, the base variants of Wav2Vec2 \cite{baevski2020wav2vec}\footnote{\scriptsize https://dl.fbaipublicfiles.com/fairseq/wav2vec/wav2vec\_small.pt}, HuBERT \cite{hsu2021hubert}\footnote{\scriptsize https://dl.fbaipublicfiles.com/hubert/hubert\_base\_ls960.pt}, and WavLM \cite{chen2022wavlm}\footnote{\scriptsize https://huggingface.co/s3prl/converted\_ckpts/resolve/main/wavlm\_base.pt}. In terms of the downstream module, we use a mean pooling layer and two linear layers. Besides, the objective function for the SER task is class-balanced cross-entropy loss \cite{Cui_2019}.

% We integrate learned features by computing a weighted sum of the SSL models' intermediate representations and mapping them to target emotion labels via a downstream linear classifier.

% \subsection{Problem Formulation}
% We address the challenge of personalized SER in a few-shot learning setting because it is impractical to collect large amounts of personalized labeled data for every user. In our framework, each task corresponds to the unique labeling behavior of an individual annotator. For a given annotator, we assume access to a small set of labeled examples for adaptation (support set) and an additional set for evaluation (query set).

% Formally, for an annotator \(i\), we define a task as \(
% \mathcal{T}_i = \{S_i, Q_i\},
% \)
% where the support set is
% \(
% S_i = \{(x_{ij}, y_{ij})\}_{j=1}^{K}
% \)
% and the query set is
% \(
% Q_i = \{(x_{ij}, y_{ij})\}_{j=1}^{Q}.
% \)

% Here, \(x_{ij}\) represents the acoustic features and \(y_{ij}\) the corresponding emotion labels provided by annotator \(i\). The objective is to rapidly adapt a generic SER model to capture the new annotator’s emotion perception using only a few examples.

\subsection{Problem Formulation}
We address the challenge of personalized SER in a few-shot learning setting ince collecting large amounts of personalized labeled data for every user is impractical in real-world scenarios. For each experiment, we divide the data into three sets: a training set, a few-shot training set, and a few-shot testing set.
The training set \(\mathcal{T}=\{\mathcal{T}_i\}_{i=1}^N\) consists of a large set of labeled examples annotated by N annotators, used for general model training. For each new annotator \(f\), we withhold their data \(\mathcal{T}_f\) from the training set. From this held-out data, we construct a few-shot training set \(\mathcal{F}_{train,f} = \{(x_{fj}, y_{fj})\}_{j=1}^{K}\), which contains \(K\) labeled speech samples used for model adaptation. Finally, the few-shot testing set \(\mathcal{F}_{test,f} = \{(x_{fj}, y_{fj})\}_{j=1}^{Q}\) consists of an additional \(Q\) labeled speech samples from annotator \(f\), used exclusively for evaluation.
Here, \(x_{fj}\) represents the speech signal and \(y_{fj}\) the corresponding emotion labels provided by annotator \(f\). The objective is to rapidly adapt a generic SER model to capture the new annotator’s emotion perception using only a few personalized examples.

\begin{table*}[t]
\centering
\fontsize{8}{9}\selectfont
\caption{\small This table summarizes the results for both "Seen Data" and "Unseen Data" scenarios using WavLM-, Wav2vec2-, and HuBERT-based SER models. All values are presented in \textbf{percentages} (\textbf{\%}), and the best performance for metrics is highlighted in \textbf{bold}.}
\vspace{-3mm}
\begin{tabular}{@{\hspace{0.2cm}}c|@{\hspace{0.2cm}}c@{\hspace{0.2cm}}c@{\hspace{0.2cm}}c|@{\hspace{0.2cm}}c@{\hspace{0.2cm}}c@{\hspace{0.2cm}}c|@{\hspace{0.2cm}}c@{\hspace{0.2cm}}c@{\hspace{0.2cm}}c|@{\hspace{0.2cm}}c@{\hspace{0.2cm}}c@{\hspace{0.2cm}}c|@{\hspace{0.2cm}}c@{\hspace{0.2cm}}c@{\hspace{0.2cm}}c|@{\hspace{0.2cm}}c@{\hspace{0.2cm}}c@{\hspace{0.2cm}}c@{\hspace{0.cm}}}
\toprule
Scenario           & \multicolumn{9}{c}{Seen Data}                                                                                                                          & \multicolumn{9}{c}{Unseen Data}                                                                                                                        \\ \midrule
Upstream          & \multicolumn{3}{c}{WavLM}                        & \multicolumn{3}{c}{Wav2vec2}                     & \multicolumn{3}{c}{HuBERT}                       & \multicolumn{3}{c}{WavLM}                        & \multicolumn{3}{c}{Wav2vec2}                     & \multicolumn{3}{c}{HuBERT}                       \\  \midrule
Metrics            & maF1           & miF1           & UA             & maF1           & miF1           & UA             & maF1           & miF1           & UA             & maF1           & miF1           & UA             & maF1           & miF1           & UA             & maF1           & miF1           & UA             \\ \midrule
Random             & 14.6          & 17.4          & 60.7          & 14.6          & 17.4          & 60.7          & 14.6          & 17.4          & 60.7          & 14.7          & 18.0          & 60.8          & 14.7          & 18.0          & 60.8          & 14.7          & 18.0          & 60.8          \\ \midrule
Linear-Few         & 21.7           & 34.5           & 64.9           & 21.9           & 35.1           & 65.4           & 21.3           & 34.0           & 64.0           & 20.4           & 33.5           & 62.9           & 19.9           & 34.1           & 63.7           & 20.0           & 32.6           & 60.1           \\
Entire-Few              & 34.3           & 48.0           & 80.7           & 34.2           & 47.5           & 79.8           & 33.9           & 47.1           & 80.0           & 24.8           & 39.7           & 75.9           & 23.8           & 39.7           & 75.0           & 25.6           & 42.3           & 77.0           \\
Entire-Zero           & 32.5           & 43.8           & 77.8           & 26.6           & 34.8           & 63.1           & 26.9           & 35.2           & 63.5           & 24.7           & 37.4           & 74.4           & 24.6           & 38.0           & 74.5           & 25.0           & 40.3           & 74.8           \\
Multi-Few         & 23.3           & 29.4           & 62.7           & 22.3           & 28.4           & 63.5           & 23.8           & 30.0           & 64.8           & 20.0           & 33.7           & 66.5           & 19.4           & 33.9           & 67.9           & 20.8           & 33.2           & 67.0           \\
Entire-Sim.              & 24.4           & 35.8           & 63.5           & 24.4           & 37.0           & 64.1           & 25.5           & 37.2           & 64.8           & 21.8           & 33.7           & 56.1           & 21.5           & 33.7           & 57.1           & 21.3           & 33.5           & 56.3           \\
\midrule  
\cellcolor{gray!20}\textbf{Meta-PerSER} & \textbf{35.7}  & \textbf{50.7}  & \textbf{82.8}  & \textbf{35.6}  & \textbf{48.7}  & \textbf{81.3}  & \textbf{35.3}  & \textbf{51.2}  & \textbf{82.4}  & \textbf{30.7}  & \textbf{47.6}  & \textbf{80.9}  & \textbf{27.2}  & \textbf{44.9}  & \textbf{78.7}  & \textbf{30.2}  & \textbf{46.4}  & \textbf{80.0}  \\ \bottomrule 
\end{tabular}
\label{tab:overall_results}
\vspace{-6mm}
\end{table*}

\subsection{Meta-PerSER Framework}
Meta-PerSER is a novel meta-learning framework designed to improve the efficiency and personalization of SER. Based on Model-Agnostic Meta-Learning (MAML), our approach integrates multiple enhancements to improve generalization, reduce computational costs, and optimize training efficiency.

\subsubsection{Pretraining for General Emotion Representation}
\label{subsubsec:ssl-e}
Before personalization, we first train a base SER model to capture general emotional patterns. This model is trained on training set \( \mathcal{T} \), which consists of labeled emotion data aggregated from multiple annotators. We denote this pretrained model as \textbf{SSL-E}, which serves as a strong initialization for subsequent adaptation.

\subsubsection{Combined-Set Meta-Training (CSMT)}
To enable rapid adaptation to individual annotators' unique emotion perceptions, we employ meta-learning. Traditional MAML splits data into support and query sets to simulate task adaptation and evaluation separately. However, in our setting, all annotators label data from the same domain and share similar feature distributions. As a result, the distinction between adaptation and evaluation is less meaningful, and splitting the data provides little benefit. Instead, we introduce \textbf{Combined-Set Meta-Training (CSMT)}, where we sample and update the model using all available training data. This approach allows the model to leverage a larger, more diverse dataset during training, leading to improved generalization across annotators without unnecessary partitioning.

\subsubsection{Efficient Gradient Optimization}
Computational efficiency is a key challenge in meta-learning, particularly due to the high cost of computing second-order derivatives in standard MAML. To address this, we apply \textbf{Derivative Annealing (DA)} \cite{antoniou2018train}, which initially uses only first-order gradients during early training and gradually introduces second-order derivatives in later stages. This approach significantly reduces computational overhead while still capturing higher-order interactions when needed.  

In addition, we integrate \textbf{Learning Per-Layer Per-Step Learning Rates and Gradient Directions (LSLR)} \cite{antoniou2018train}, a dynamic learning rate adjustment mechanism that optimizes per-layer learning rates at each update step. While DA improves computational efficiency by mitigating second-order derivative costs, LSLR enhances adaptation stability and reduces the need for extensive hyperparameter tuning.

\subsubsection{Meta-PerSER Training and Adaptation Process}
The overall Meta-PerSER workflow consists of two phases: meta-training and meta-testing.

\textbf{Meta-Training Phase:}  
The meta-training process begins with initializing SSL-E, which serves as the starting point for adaptation. In the inner loop, for each annotator $i$ we randomly sample \(K\) examples from \(\mathcal{T}_i\) as \(\mathcal{T}_i^\prime\), compute the inner loss \(\mathcal{L}(\theta, \mathcal{T}_i^\prime)\), and update model weights using LSLR to obtain annotator-specific parameters \(\theta_i\). Following this, the outer loop aggregates gradients from multiple annotators using $\mathcal{L}(\theta_i, \mathcal{T}_i^\prime)$, updating shared initialization \(\theta\) to improve generalization. To optimize computational efficiency, we apply Derivative Annealing (DA), which restricts gradient computation to first-order derivatives in the early stages (first 30\% of inner-loop steps) before transitioning to second-order derivatives later for richer adaptation. 

\textbf{Meta-Testing Phase:}  
During the meta-testing phase, a new, previously unseen annotator is introduced to assess the model's adaptability. The model undergoes adaptation using $K$ samples from annotator $f$,  \(\mathcal{F}_{train,f}\). Finally, its performance is evaluated on $Q$ samples in \(\mathcal{F}_{test,f}\), thereby testing its generalization capability on novel annotator data.

This two-phase process ensures that Meta-PerSER enables rapid adaptation to new annotators while maintaining strong generalization.

\vspace{-3pt}

\subsection{Baselines}
\vspace{-1mm}
Our baselines are named according to a two-stage process: first, how the model initialization is trained before few-shot adaptation, followed by a hyphen (``-"), and second, how the few-shot training set $\mathcal{F}_{train}$ is employed for adaptation.

\textbf{SSL Approach:} We train SSL-based SER models using aggregated labels from training annotators, resulting in the SSL-E model in Sec.~\ref{subsubsec:ssl-e}. We then adapt them to unseen annotators during meta-testing. We evaluated three strategies: the \textbf{Entire-Few} baseline, where the SSL-E model is fine-tuned with few-shot training set \(\mathcal{F}_{train}\) during meta-testing; \textbf{Linear-Few}, in which only the classifier and weighted sum of SSL features are trained on the entire training set prior to few-shot adaptation; and \textbf{Entire-Zero}, where the SSL-E model is evaluated in a zero-shot manner on the few-shot testing set \(\mathcal{F}_{test}\) without any further fine-tuning.

\textbf{Multi-Task Approach (Multi-Few)}: We models each annotator’s labeling behavior through individual classification tasks. While the upstream model is shared, each annotator has a separate prediction head in the classifier. After training, meta-testing is used to learn a new classifier from random initialization for unseen annotators.

\textbf{Similarity-based Approach (\textbf{Entire-Sim.})}
\cite{shi2020few}: We use features extracted by the SSL-E model to classify $\mathcal{F}_{test}$ samples based on their proximity to few-shot training samples in the feature space. For each emotion $e$, we compute a center $c_e$ by averaging the SSL feature embeddings of all few-shot training samples labeled with that emotion. Then, for each few-shot testing sample $\bm{x}$, we calculate the cosine similarity $s$ between its SSL feature embedding $f(x)$ and each emotion center $c_e$. The similarity is transformed via a softmax function to produce a probability distribution over the emotion labels. The probability that $\bm{x}$ is assigned to the emotion label $e_i$ is:
\begin{equation}
    p(y=e_i|x)=\frac{exp(s(c_{e_i}, f(x)))}{\sum_{e\in\mathcal{E}} exp(s(c_{e}, f(x)))},
\end{equation}
where $\mathcal{E}$ is the set of all emotions.

% \subsubsection{Meta-learning (Meta) Approach}
% { \color{red} Need more descriptions.

% We utilize the standard meta-learning approach (\textbf{Meta} in Table \ref{tab:overall_results}) as another baseline for comparison in the work.
% }
\vspace{-3pt}

\section{Experiments Settings}
\vspace{-1mm}
% We design four experiments to evaluate our proposed framework. First, an ablation study is conducted to demonstrate the effectiveness of the three techniques incorporated into our method: omitting the support-query set split, derivative annealing, and LSLR. The second experiment investigates the optimal support set size that balances performance and computational cost. The final two experiments evaluate our approach and the baseline models under different dataset splitting strategies. The first strategy, ``without session split,'' examines model performance when encountering new annotators with previously seen audio samples. The second strategy, ``session split,'' assesses the model’s ability to generalize to new annotators with entirely unseen audio samples.

% \subsection{Model architecture}

% We adopt a unified model architecture across all experiments by following the EMO-SUPERB \cite{wu2024emo} based on s3prl toolkit \cite{Yang_2021}. Specifically, the SER framework has two main parts. The first part is upstream, meaning the ptr-trained self-supervised learning (SSL) models. Then, the downstream is to train the model for SER task. We utilize the base variants of WavLM \cite{chen2022wavlm}, Wav2Vec2 \cite{baevski2020wav2vec}, and HuBERT \cite{hsu2021hubert} as upstream models and  compute a weighted sum of the SSL model's intermediate representations to effectively integrate the learned features with downstream tasks, employing linear classifier to map these aggregated features to the target emotion labels.

\subsection{Resource}
We use the IEMOCAP corpus \cite{busso2008iemocap}, which contains 10,039 utterances by 10 professional actors. Each utterance is labeled by at least three annotators, who may assign one or more emotion labels from a set of 10 categories—namely, frustrated, angry, sad, disgust, excited, fear, neutral, surprise, happy, and “other.” We exclude the “other” category, defining a 9-class SER task. In total, 12 annotators participated in the labeling process. One annotator was excluded due to insufficient data, leaving 11, comprising 6 self-reporting actors and 5 external evaluators. For testing, we use only the 5 external annotators, as they each provided annotations across all 5 sessions. In each test run, one of 5 external annotators is used for testing, one for validation, and the remaining nine for training. %In meta-training, we use $N=9$ training annotators. % , and we focus on five annotators who provided ample data for developing and evaluating personalized SER models. %Furthermore, the five recruited workers are considered the primary subjects for developing personalized SER models because of the number of labeled data, which are evaluated under two distinct scenarios in the subsequent section.

% The IEMOCAP corpus \cite{busso2008iemocap} contains human-annotated 10,039 utterances from 10 professional actors. Each utterance is labeled by three annotators, allowed to choose one or more emotions from 10 emotions, including frustrated, angry, sad, disgust, excited, fear, neutral, surprise, happy, and ``other'' emotions. We exclude ``other'' to define the SER task as 9-class SER task. The number of annotators is 12. We regard the 6 recruited workers as our main targets to build the personalized SER models in two different scenarios, introduced in the next section.
\vspace{-4pt}
\subsection{Application Scenarios}
% The objective of this study is to develop personalized SER models that address inherent biases arising from factors such as varied past emotional experiences and gender differences \cite{Cowen_2017}. 
To validate the efficacy of the proposed methods, we propose two distinct application scenarios.

% Our goal is to build personalized SER models because of natural bias, such as different past emotional experiences or genders. We define two application scenarios to validate the effectiveness of the proposed methods.

% \subsubsection{Seen Data}
% The first scenario is Seen Data, indicating the trained SER systems might have seen the utterance during training, and we can use some of annotated labels on the seen data to do personalization on new annotators. To simulate the condition, we split the dataset based on annotators. The original IEMOCAP dataset includes 12 annotators, and we removed one annotator due to the number of labeled data is insufficient, leaving 11 remaining annotators. Among them, five annotators provided a sufficient number of annotations, larger than 64. To explore the robustness of methods, we run five independent experiments, each time selecting two of these five annotators as validation and test sets, while using the remaining annotators for training. The final results were obtained by averaging across all five runs. This splitting method evaluates whether the model can distinguish variations in annotation styles when exposed to the same audio.

\subsubsection{Seen Data}
In the Seen Data scenario, the SER system has already encountered the utterances during training but has not seen the labels provided by unseen annotators. %This setting allows us to use the annotated labels on these utterances to personalize the model for new annotators. 
To simulate this, we partitioned the IEMOCAP dataset by annotator. 
%Out of the original 12 annotators, we excluded one due to insufficient data, leaving 11. 
% Among these, five annotators provided over 200 annotations each, which we consider sufficient for robust evaluation. 
We conducted five independent experiments; in each, one of these five annotators was designated for testing, while the remaining annotators were used for training. The final results are averaged across all experiments, assessing whether the model can effectively capture variations in annotation styles for the same audio.

% The first scenario is Seen Data, indicating the trained SER systems might have seen the data during training, and we can use some of annotated labels on the seen data to do personalization on new annotators. To simulate the condition, we split the dataset based on annotators. The original IEMOCAP dataset includes 12 annotators, and we removed one annotator due to the number of labeled data is insufficient, leaving 11 remaining annotators. Among them, five annotators provided a sufficient number of annotations, larger than 64. To explore the robustness of methods, we run five independent experiments, each time selecting two of these five annotators as validation and test sets, while using the remaining annotators for training. The final results were obtained by averaging across all five runs. This splitting method evaluates whether the model can distinguish variations in annotation styles when exposed to the same audio.
\vspace{-3pt}
\subsubsection{Unseen Data}
%In the Unseen Data scenario, we ensure that no utterances overlap across the meta-training and meta-testing datasets. To achieve this, the predefined five-session division of the IEMOCAP dataset is employed, guaranteeing that sessions do not overlap.
In the Unseen Data scenario, we use the IEMOCAP dataset's predefined five-session split to prevent utterance overlap between meta-training and meta-testing. 
%One annotator is chosen for validation, 
One annotator is chosen for testing, and the rest for training; sessions 1–4 are allocated for training, and session 5 for testing. The final training and test sets were composed based on the annotations provided by the selected annotators within each session. This ensures no duplicate audio samples across splits, though it reduces the dataset size. % This approach prevents the model from being exposed to identical audio samples during training, albeit at the expense of a reduced dataset size.

% In the unseen data scenario, we make sure the utterance are not overlapped in the train, development, and test sets. To achieve this goal, we utilize the pre-defined five-session division in the IEMOCAP, ensuring no overlap between sessions. First, we selected one annotator for validation, one for testing, and the rest for training. Then, we selected three sessions (session 1-3) for training, one (session 4) for validation, and one (session 5) for testing. The final training, validation, and test sets were determined by the annotations provided by the selected annotators within each session. This method prevents the model from being exposed to the same audio during training, but at the cost of a reduced dataset size.

\begin{figure}[t]  
    \centering
    \vspace{-1mm}
    \includegraphics[width=0.49\textwidth]{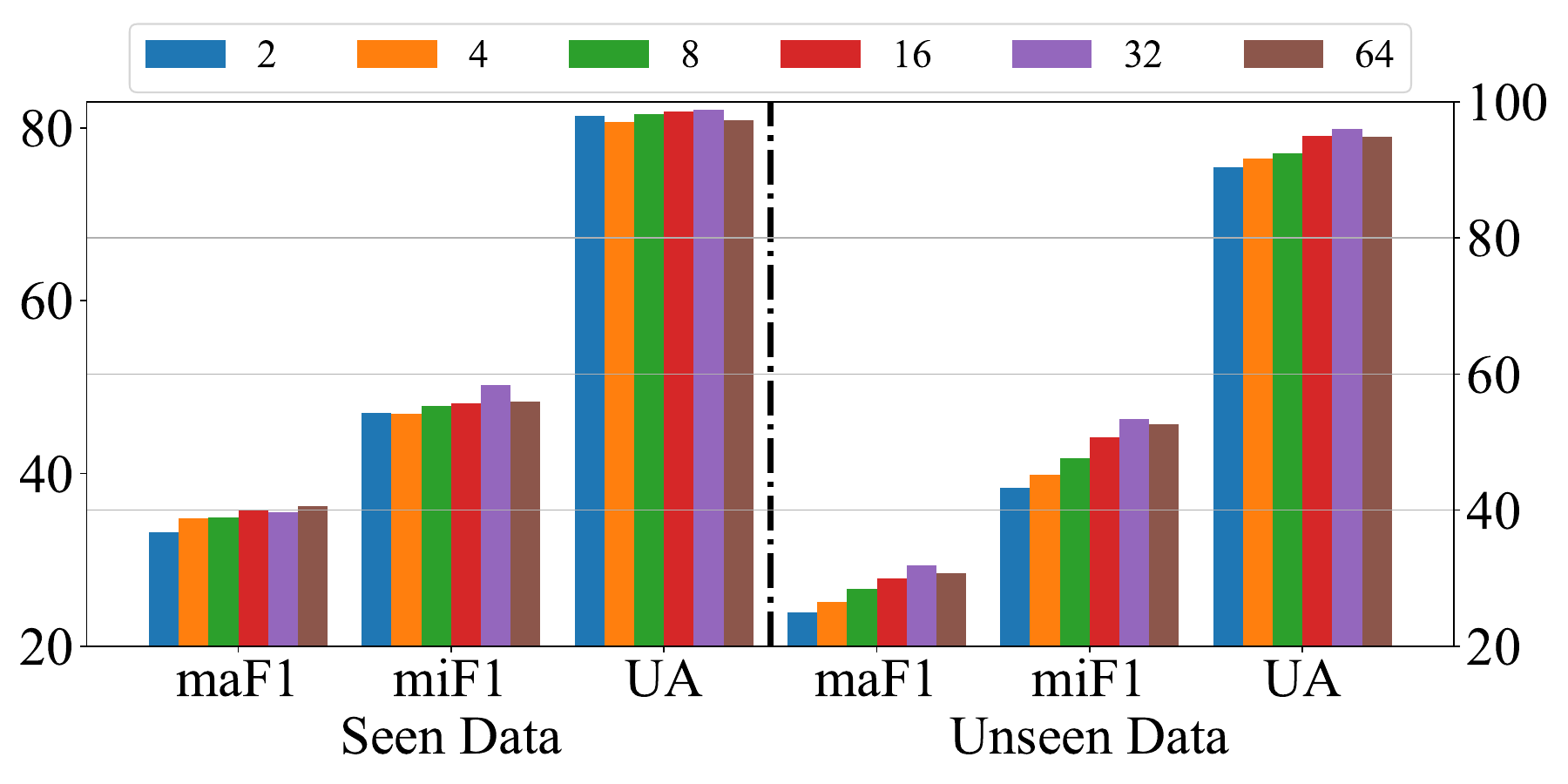} 
    \vspace{-8mm}
    \caption{Illustration of the averaged results across different numbers of few-shot training samples (shown on legend) under two scenarios in terms of macro-F1 (\textbf{maF1}), micro-F1 (\textbf{miF1}) and unweighted accuracy (\textbf{UA}) in \textbf{percentages} (\textbf{\%}).}
    \label{fig:support_size} 
    \vspace{-6mm}
\end{figure}
\vspace{-4pt}
\subsection{Evaluation}
 We conduct few-shot meta-testing using data annotated by a single annotator. To ensure robustness, each experiment is repeated 10 times with different random seeds. In each iteration, we randomly select \(K = 32\) samples for few-shot training and \(Q = 128\) samples for few-shot testing. We update the model with one batch and 50 steps.
 
The SER task is defined as a multi-label classification problem in this study, diverging from conventional approaches, to accommodate the inherent ambiguity of emotion perception. All annotators are permitted to select one or more emotions from the provided options. Following the procedure described in \cite{Wu_2024}, we convert the model's prediction probabilities into binary outcomes by applying a threshold of \(1/|\mathcal{Y}|\), where \(\mathcal{Y}\) represents the complete set of emotion classes. We employ macro-F1 (\textbf{maF1}) and micro-F1 (\textbf{miF1}) scores \cite{opitz2019macro, lin24i_interspeech} along with unweighted accuracy (UA) as evaluation metrics for assessing and reporting multi-label classification performance. % Macro-F1 provides an average of F1 scores across all emotion classes, ensuring balanced evaluation even when class frequencies vary; micro-F1 aggregates predictions to reflect overall performance; and UA offers a straightforward measure of accuracy without favoring more common labels.

% We define the SER task as multi-label task, different from the conventional previous works, to embrace ambiguity of emotion perception. All annotators are allowed to choose one or more emotions from the given options, so we choose variants of F1 scores (macro- and micro-F1 scores) \cite{opitz2019macro} unweighted accuracy (UA) as our metrics for evaluating and reporting multi-label classification performance. 

% \input{tables/support_size}

\vspace{-5pt}
\section{Results and Analysis}
\subsection{Proposed Meta-PerSER}
% Table~\ref{tab:overall_results} presents the averaged results of different approaches under Seen Data and Unseen Data scenarios. From these tables, we observe that Meta-PerSER consistently outperforms all baseline methods in both settings, demonstrating its ability to effectively adapt to unseen annotators. 

Table~\ref{tab:overall_results} demonstrates that Meta-PerSER consistently outperforms all baseline methods across both Seen and Unseen Data scenarios and across all upstream models. Under the Seen Data setting, Meta-PerSER improves macro-F1, micro-F1, and unweighted accuracy scores by approximately 1–2 percentage points over the best SSL fine-tuning baseline (SSL-FT), indicating enhanced adaptation even when the model has previously encountered the audio. In the more challenging Unseen Data scenario, where the model must generalize to entirely new audio samples, Meta-PerSER shows a larger performance margin, with improvements of up to 7 percentage points in key metrics compared to SSL-based methods.

These results validate that our integration of CSMT, DA, and LSLR effectively enhances the model's ability to quickly adapt to unseen annotators while preserving robust emotion representations, surpassing both traditional fine-tuning and multi-task strategies.

% Additionally, by comparing the two tables, we find that the improvement in F1-score from SSL to Meta-PerSER is more pronounced in Table~\ref{table:session_split} than in Table~\ref{table:without_session_split}. This indicates that Meta-PerSER is capable of successfully adapting to new annotators even when encountering previously unseen audio samples, whereas other baseline methods fail to achieve comparable performance.
\vspace{-5pt}
\subsection{Impacts of Few-Shot Training Set Size}
Figure~\ref{fig:support_size} illustrates the average performance of Meta-PerSER across different models using various few-shot training set sizes (the original results are summarized in Append Table \ref{tab:detailed_support_size}). 
% A support size of 32 yields the best overall performance in this study. However, under the Seen Data scenario, increasing the support size leads to further improvements in macro-F1 scores.
Overall, performance improves as the few-shot training set size increases from 2 to 32 samples. 
For seen data, macro-F1 increases from 33.2 to 35.5, micro-F1 reaches a peak of 50.2, and unweighted accuracy (UA) improves to 82.1 at 32 samples.
Similarly, for unseen data, all metrics steadily rise up to a few-shot training set size of 32. 
However, increasing the few-shot training set size to 64 results in a slight decline in performance, particularly for unseen data. 
This suggests that while a larger few-shot training set initially provides more representative examples for effective adaptation, exceeding an optimal size (in this case, 32 samples) may introduce outlier or ambiguous samples that confuse the model.
\vspace{-5pt}

\subsection{Effectiveness of Different Techniques}
Table \ref{tab:ablation} presents the ablation study on techniques, proving the four components of Meta-PerSER collectively enhance performance under the Unseen Data scenario. Specifically, the incorporation of \textbf{INI} (i.e., leveraging the pre-trained \textbf{SSL-E} weights) yields a significant improvement in the macro-F1 metric across all three SSL-based models. Additionally, the individual application of CSMT, DA, and LSLR consistently enhances performance across all evaluation metrics. The combination of
\textbf{CSMT} and \textbf{INI} generally outperforms configurations using either (\textbf{DA} 
and \textbf{INI}) or (\textbf{LSLR} and \textbf{INI}) in most cases. Most importantly, the comprehensive integration of all three methods produces the highest gains in macro-F1, micro-F1, and unweighted accuracy across all upstream models. These findings robustly validate that our integrated approach, \textbf{Meta-PerSER}, substantially augments model in unseen scenarios, thereby advancing the state-of-the-art for personalized SER.

% The ablation study in Table~\ref{tab:ablation} shows that four components of proposed Meta-PerSER contributes to improved performance under Unseen Data scenario. The model with the \textbf{INI} (using model weights of the \textbf{SSL-E}) can improve macro-F1 metric across three SSL-based models. 
% Then, using CSMT, DA, and LSLR individually enhances all three metrics. The model using \textbf{CSMT} and \textbf{INI} improves the SER performance, comparing to model with \textbf{DA} and the model with \textbf{LSLR}.
% Most importantly, integrating all three methods produces the highest improvements across macro-F1, micro-F1, and unweighted accuracy for all upstream models. 
% These findings confirm that our combined approach significantly boosts the model's generalization and adaptation to unseen data, validating the design of Meta-PerSER.

\begin{table}[t]
\centering
\fontsize{7}{9}\selectfont
\caption{Table summarizes the performance of adding the proposed components one by one in \textbf{percentages} (\textbf{\%}). \textbf{INI} indicates whether the model initialize from the pre-trained \textbf{SSL-E}; \textbf{\checkmark} indicates that the model incorporates the corresponding techniques. \textbf{CSMT}, \textbf{DA}, and \textbf{LSLR} are defined in Section~\ref{s:method}.}
% \begin{tabular}{@{}ccccccccccccc@{}}
\vspace{-3mm}
\begin{tabular}{@{\hspace{0.05cm}}c@{\hspace{0.05cm}}c@{\hspace{0.05cm}}c@{\hspace{0.05cm}}c|@{\hspace{0.05cm}}c@{\hspace{0.05cm}}c@{\hspace{0.05cm}}c|@{\hspace{0.05cm}}c@{\hspace{0.05cm}}c@{\hspace{0.05cm}}c|@{\hspace{0.05cm}}c@{\hspace{0.05cm}}c@{\hspace{0.05cm}}c@{\hspace{0.05cm}}}
\toprule
% \multicolumn{4}{c}{Scenario} & \multicolumn{9}{c}{Unseen Data}                                                                                                               \\ \midrule
\multicolumn{4}{c}{Upstream} & \multicolumn{3}{c}{WavLM}                     & \multicolumn{3}{c}{Wav2vec2}                  & \multicolumn{3}{c}{HuBERT}                    \\ \toprule
\textbf{INI}   & \textbf{CSMT}   & \textbf{DA}   & \textbf{LSLR}  & maF1          & miF1          & UA            & maF1          & miF1          & UA            & maF1          & miF1          & UA            \\ \midrule
     &        &      &       & 22.8          & 40.1          & 76.8          & 23.6          & 40.9          & 77.4          & 23.6          & 40.7          & 78.1          \\ \midrule
\checkmark    &        &      &       & 25.7          & 39.3          & 74.8          & 24.9          & 39.4          & 72.2          & 27.1          & 41.9          & 76.3          \\
\checkmark & \checkmark      &      &       & 27.6          & 40.7          & 77.3          & 26.6          & 39.1          & 74.5          & 28.1          & 41.5          & 77.5          \\
\checkmark &        & \checkmark    &       & 27.9          & 41.8          & 77.0          & 24.9          & 39.4          & 72.2          & 27.0          & 41.7          & 76.3          \\
\checkmark &        &      & \checkmark     & 27.2          & 43.8          & 77.1          & 25.3          & 42.4          & 75.4          & 26.8          & 44.5          & 78.1          \\ \midrule
\checkmark     & \checkmark      & \checkmark    & \checkmark     & \textbf{30.7} & \textbf{47.6} & \textbf{80.9} & \textbf{27.2} & \textbf{44.9} & \textbf{78.7} & \textbf{30.2} & \textbf{46.4} & \textbf{80.0} \\ \bottomrule
\end{tabular}
\label{tab:ablation}
\vspace{-8mm}
\end{table}

\vspace{-5pt}
\section{Limitations}
We are among the first to investigate categorical personalized SER systems. However, our current experimental settings do not incorporate conversational context, as emotional ratings are based solely on conversational-level audio-visual cues provided by annotators. This limitation may hinder the system’s ability to accurately capture the natural context of individual emotion perception.

\vspace{-5pt}
\section{Conclusion and Future Work}
This paper introduces a novel framework, \textbf{Meta-PerSER}, designed to effectively adapt to unseen annotators in SER tasks with only a few labeled examples. Meta-PerSER integrates a pre-trained self-supervised backbone with Combined-Set Meta-Training, Derivative Annealing, and per-layer adaptive learning rates. This design enables our system to outperform traditional fine-tuning and multi-task baselines in both seen and unseen scenarios on the IEMOCAP corpus. Our results demonstrate that meta-learning can capture subjective emotion perceptions without extensive annotation collection.

While previous research in SER has primarily focused on speaker personalization, this work shifts the focus toward listener personalization—specifically, adapting to previously unseen annotators. For future research, we plan to explore the applicability of Meta-PerSER to other subjective learning tasks beyond SER, such as hate speech detection and customer experience recognition. We also plan to investigate alternative meta-adaptation strategies and extend our approach to multilingual and low-resource settings.

% \clearpage
\bibliographystyle{IEEEtran}
\bibliography{mybib}

\clearpage
\small
\appendix
\renewcommand{\thetable}{A\arabic{table}} % 設定 Table 標號為 A1, A2...
\renewcommand{\thefigure}{A\arabic{figure}} % 設定 Figure 標號為 A1, A2...
\setcounter{table}{0}  % 附錄中的表格重新從 1 開始計數
\setcounter{figure}{0} % 附錄中的圖重新從 1 開始計數

\begin{center}
    \Large\bfseries Supplementary Material 
\end{center}

In this Supplementary Material, we provide additional experimental results, detailed emotional annotation distributions, and further training details to facilitate reproducibility.

\section{Additional Experimental Results}
Table~\ref{tab:detailed_support_size} presents the performance across various numbers of shots (2, 4, 8, 16, 32, and 64) under two scenarios, evaluated in terms of macro-F1 (\textbf{maF1}), micro-F1 (\textbf{miF1}), and unweighted accuracy (\textbf{UA}). Models employing 32 shots achieve the best overall performance.

\section{Rater-based Annotation Distribution}
To highlight annotator differences, we summarize the distribution of emotion classes for each annotator in the test set. Table \ref{tab:seen_dist} presents the results for the ``Seen Data" scenario, while Table \ref{tab:unseen_dist} shows those for the ``Unseen Data" scenario. Each element represents the proportion of a specific emotion's annotations relative to the total annotations made by that annotator, and the final row displays the average number of annotations per data point.

% To highlight the differences between annotators, we summarize the emotional distribution corresponding to emotion classes for each annotator in the test set. Table \ref{tab:seen_dist} presents the results for the ``Seen Data'' scenario, while Table \ref{tab:unseen_dist} shows the results for the ``Unseen data'' scenario. Each element in these tables represents the proportion of a specific emotion's annotations relative to the total annotations made by that annotator. The last row shows the average number of annotations per data point made by each annotator.

\begin{table}[h]
     \centering
    \fontsize{8}{9}\selectfont
    \caption{Emotional annotation distribution for five annotators for evaluation under ``Seen Data'' scenario. All values are presented in \textbf{percentages} (\textbf{\%}).}
    %\begin{tabular}{lccccc}  % @{\hspace{0.2cm}}l
    \begin{tabular}{@{\hspace{0.2cm}}l@{\hspace{0.2cm}}r@{\hspace{0.2cm}}r@{\hspace{0.2cm}}r@{\hspace{0.2cm}}r@{\hspace{0.2cm}}r@{\hspace{0.2cm}}} 
        \toprule
        Annotator & C-E1 & C-E2 & C-E4 & C-E5 & C-E6 \\
        % & Chloe Menon & James Bonaiuto & Naheed Fakoor & Osaretin Edobor & Saki Matsumoto \\
        \midrule
        Frustration & 38.16 & 12.12 & 24.66 & 16.45 & 30.61 \\
        Anger & 13.21 & 22.83 & 9.57 & 11.46 & 9.87 \\
        Sadness & 12.73 & 17.53 & 8.98 & 4.07 & 9.37 \\
        Disgust & 0.13 & 0.38 & 0.08 & 5.73 & 1.01 \\
        Excited & 24.07 & 7.01 & 11.53 & 0.55 & 16.21 \\
        Fear & 1.68 & 0.45 & 0.48 & 6.10 & 1.51 \\
        Neutral & 5.02 & 17.35 & 36.78 & 48.43 & 13.09 \\
        Surprise & 0.20 & 2.85 & 1.679 & 1.48 & 4.43 \\
        Happiness & 4.79 & 19.48 & 6.24 & 5.73 & 13.90 \\
        \midrule
        \# of annotations/data & 1.02 & 1.01 & 1.02 & 1.12 & 1.04 \\
        \bottomrule
    \end{tabular}
    \label{tab:seen_dist}
\end{table}

% \begin{table}[h]
%     \centering
%     \renewcommand{\arraystretch}{1.2} % Adjust row height
%     \begin{tabular}{lcccccc} 
%         \toprule
%         & C-E1 & C-E2 & C-E4 & C-E5 & C-E6 \\
%         & Chloe Menon & James Bonaiuto & Naheed Fakoor & Osaretin Edobor & Saki Matsumoto \\
%         \midrule
%         Frustration & 0.3816 & 0.1212 & 0.2466 & 0.1645 & 0.3061 \\
%         Anger & 0.1321 & 0.2283 & 0.0957 & 0.1146 & 0.0987 \\
%         Sadness & 0.1273 & 0.1753 & 0.0898 & 0.0407 & 0.0937 \\
%         Disgust & 0.0013 & 0.0038 & 0.0008 & 0.0573 & 0.0101 \\
%         Excited & 0.2407 & 0.0701 & 0.1153 & 0.0055 & 0.1621 \\
%         Fear & 0.0168 & 0.0045 & 0.0048 & 0.0610 & 0.0151 \\
%         Neutral & 0.0502 & 0.1735 & 0.3678 & 0.4843 & 0.1309 \\
%         Surprise & 0.0020 & 0.0285 & 0.01679 & 0.0148 & 0.0443 \\
%         Happiness & 0.0479 & 0.1948 & 0.0624 & 0.0573 & 0.1390 \\
%         \midrule
%         \#annotation/\#data & 1.0216 & 1.0063 & 1.0234 & 1.1155 & 1.0365 \\
%         \bottomrule
%     \end{tabular}
%     \caption{Emotion distribution for each test annotator  in seen data scenario}
%     \label{tab:seen_dist}
% \end{table}
\begin{table}[h]
    \centering
    \fontsize{8}{9}\selectfont
    \caption{Emotional annotation distribution for five annotators for evaluation under ``Unseen Data'' scenario. All values are presented in \textbf{percentages} (\textbf{\%}).}
    %\begin{tabular}{lccccc} % 
    \begin{tabular}{@{\hspace{0.2cm}}l@{\hspace{0.2cm}}r@{\hspace{0.2cm}}r@{\hspace{0.2cm}}r@{\hspace{0.2cm}}r@{\hspace{0.2cm}}r@{\hspace{0.2cm}}}
        \toprule
        Annotator & C-E1 & C-E2 & C-E4 & C-E5 & C-E6 \\
        % & Chloe Menon & James Bonaiuto & Naheed Fakoor & Osaretin Edobor & Saki Matsumoto \\
        \midrule
        Frustration & 40.13 & 12.44 & 26.65 & 8.16 & 14.48 \\
        Anger & 10.33 & 21.43 & 6.85 & 6.12 & 3.45 \\
        Sadness & 14.72 & 17.18 & 9.33 & 13.27 & 0.00 \\
        Disgust & 0.12 & 0.37 & 0.06 & 9.18 & 0.00 \\
        Excited & 22.06 & 7.51 & 13.71 & 2.04 & 37.93 \\
        Fear & 1.62 & 0.43 & 1.52 & 27.55 & 0.00 \\
        Neutral & 5.43 & 17.92 & 35.28 & 33.67 & 22.76 \\
        Surprise & 0.23 & 2.83 & 1.14 & 0.00 & 2.76 \\
        Happiness & 5.37 & 19.89 & 5.46 & 0.00 & 18.62 \\
        \midrule
        \# of annotations/data & 1.02 & 1.01 & 1.01 & 1.21 & 1.09 \\
        \bottomrule
    \end{tabular}
    \label{tab:unseen_dist}
\end{table}

% \begin{table}[h]
%     \centering
%     \fontsize{7}{9}\selectfont
%     % \renewcommand{\arraystretch}{1.2} % Adjust row height
%     \begin{tabular}{lccccc} 
%         \toprule
%         & C-E1 & C-E2 & C-E4 & C-E5 & C-E6 \\
%         & Chloe Menon & James Bonaiuto & Naheed Fakoor & Osaretin Edobor & Saki Matsumoto \\
%         \midrule
%         Frustration & 0.4013 & 0.1244 & 0.2665 & 0.0816 & 0.1448 \\
%         Anger & 0.1033 & 0.2143 & 0.0685 & 0.0612 & 0.0345 \\
%         Sadness & 0.1472 & 0.1718 & 0.0933 & 0.1327 & 0.0000 \\
%         Disgust & 0.0012 & 0.0037 & 0.0006 & 0.0918 & 0.0000 \\
%         Excited & 0.2206 & 0.0751 & 0.1371 & 0.0204 & 0.3793 \\
%         Fear & 0.0162 & 0.0043 & 0.0152 & 0.2755 & 0.0000 \\
%         Neutral & 0.0543 & 0.1792 & 0.3528 & 0.3367 & 0.2276 \\
%         Surprise & 0.0023 & 0.0283 & 0.0114 & 0.0000 & 0.0276 \\
%         Happiness & 0.0537 & 0.1989 & 0.0546 & 0.0000 & 0.1862 \\
%         \midrule
%         \#annotation/\#data & 1.0224 & 1.0068 & 1.0096 & 1.2099 & 1.0902 \\
%         \bottomrule
%     \end{tabular}
%     \caption{Emotion distribution for each test annotator in unseen data scenario}
%     \label{tab:unseen_dist}
% \end{table}

\begin{table}[h]
    \caption{Average runtime for different phases}
    \centering
    \renewcommand{\arraystretch}{1.2} % Adjust row height
    \begin{tabular}{lccc} 
        \toprule
        & SSL-E & Meta Training & Meta Testing \\
        \midrule
        Runtime & 2.5 hours & 4 hours & 1 min \\
        \bottomrule
    \end{tabular}

    \label{tab:runtime_comparison}
\end{table}

\begin{table*}[htbp!]
\centering
\fontsize{8}{9}\selectfont
\caption{\small Illustration of the results across different support sizes under two scenarios in terms of macro-F1 (\textbf{maF1}), micro-F1 (\textbf{miF1}) and unweighted accuracy (\textbf{UA}).}
% \begin{tabular}{@{}c|ccc|ccc|ccc|ccc|ccc|ccc@{}}
\vspace{-3mm}
\begin{tabular}{@{\hspace{0.2cm}}c@{\hspace{0.2cm}}|@{\hspace{0.2cm}}c@{\hspace{0.2cm}}c@{\hspace{0.2cm}}c@{\hspace{0.2cm}}|@{\hspace{0.2cm}}c@{\hspace{0.2cm}}c@{\hspace{0.2cm}}c@{\hspace{0.2cm}}|@{\hspace{0.2cm}}c@{\hspace{0.2cm}}c@{\hspace{0.2cm}}c@{\hspace{0.2cm}}|@{\hspace{0.2cm}}c@{\hspace{0.2cm}}c@{\hspace{0.2cm}}c@{\hspace{0.2cm}}|c@{\hspace{0.2cm}}c@{\hspace{0.2cm}}c@{\hspace{0.2cm}}|c@{\hspace{0.2cm}}c@{\hspace{0.2cm}}c@{\hspace{0.2cm}}}
\toprule
Scenario  & \multicolumn{9}{c|}{Seen Data}                                                                                                                 & \multicolumn{9}{c}{Unseen Data}                                                                                                               \\ \midrule
Upstream & \multicolumn{3}{c}{WavLM}                     & \multicolumn{3}{c}{Wav2vec2}                  & \multicolumn{3}{c}{HuBERT}                    & \multicolumn{3}{c}{WavLM}                     & \multicolumn{3}{c}{Wav2vec2}                  & \multicolumn{3}{c}{HuBERT}                    \\ \midrule
\# of Shots   & maF1          & miF1          & UA            & maF1          & miF1          & UA            & maF1          & miF1          & UA            & maF1          & miF1          & UA            & maF1          & miF1          & UA            & maF1          & miF1          & UA            \\ \midrule
2         & 32.7          & 47.6          & 81.7          & 34.8          & 47.3          & 80.8          & 32.2          & 46.0          & 81.5          & 24.6          & 40.6          & 77.5          & 22.1          & 35.8          & 73.3          & 25.0          & 38.5          & 75.5          \\
4         & 34.4          & 47.1          & 80.8          & 35.1          & 46.9          & 80.6          & 34.9          & 46.6          & 80.8          & 26.7          & 42.7          & 78.0          & 23.3          & 37.3          & 74.2          & 25.3          & 39.4          & 76.8          \\
8         & 35.5          & 48.9          & 82.1          & 34.7          & 47.2          & 81.1          & 34.6          & 47.3          & 81.5          & 28.0          & 44.8          & 79.5          & 24.4          & 38.7          & 74.6          & 27.4          & 41.9          & 77.2          \\
16        & 35.6          & 49.3          & \textbf{83.1} & 35.1          & 46.8          & 80.9          & 36.2          & 48.1          & 81.7          & 28.7          & 44.9          & 79.2          & 26.7          & 43.4          & 78.3          & 28.1          & 44.4          & 79.9          \\
32        & 35.7          & \textbf{50.7} & 82.8          & 35.6          & \textbf{48.7} & \textbf{81.3} & 35.3          & \textbf{51.2} & \textbf{82.4} & \textbf{30.7} & \textbf{47.6} & \textbf{80.9} & \textbf{27.2} & \textbf{44.9} & \textbf{78.7} & \textbf{30.2} & \textbf{46.4} & \textbf{80.0} \\
64        & \textbf{36.0} & 48.6          & 81.2          & \textbf{36.5} & 48.6          & 81.1          & \textbf{36.0} & 47.6          & 80.5          & 29.1          & 45.8          & 78.9          & 26.9          & 44.9          & 78.5          & 29.2          & 46.3          & 79.7          \\ \bottomrule
\end{tabular}
\label{tab:detailed_support_size}
\end{table*}

\section{Training Details for Reproducibility}
During training, we select one annotator from the training set as the validation annotator. We fix each testing annotator to have a corresponding validation annotator. We leave out the validation annotator during meta-training and perform meta-testing on the validation annotator to select the best model among all meta-training steps.

We employ the AdamW optimizer with a learning rate of 0.00009 in the outer loop and a batch size of 32. In the inner loop, the learning rate is initialized at 0.001. The best-performing models are selected based on the lowest class-balanced cross-entropy loss on the development set. All experiments are conducted on Nvidia Tesla V100 GPUs (32 GB), with a total computational cost of approximately 1,300 GPU hours, the detailed runtime for each baseline are listed in Table \ref{tab:runtime_comparison}. The number of parameters in our models is 94M.

% We use the AdamW optimizer with a 0.00009 learning rate and the batch size is 32. In the inner loop, we set 0.001 as our initial learning rate. We choose the best models according to the lowest value of the class-balanced cross-entropy loss on the development set. We use the Nvidia Tesla v100 GPUs with 32 GB memory and Nvida GeForce RTX 3090 with 24 GB memory for all esults. The total of GPU hours is around 1,300 hours.

\end{document}